# The effect of messaging and gender on intentions to wear a face covering to slow down COVID-19 transmission

Valerio Capraro[1] and Hélène Barcelo[2]

[1] Middlesex University London, UK
[2] Mathematical Science Research Institute, Berkeley, USA

Contact author: v.capraro@mdx.ac.uk

**Abstract**

Now that various countries are or will soon be moving towards relaxing shelter-in-place rules, it is important that people use a face covering, to avoid an exponential resurgence of the spreading of the coronavirus disease (COVID-19). Adherence to this measure will be made explicitly compulsory in many places. However, since it is impossible to control each and every person in a country, it is important to complement governmental laws with behavioral interventions devised to impact people's behavior beyond the force of law. Here we report a pre-registered online experiment (N=2,459) using a heterogenous, although not representative, sample of people living in the USA, where we test the relative effect of messages highlighting that the coronavirus is a threat to "you" vs "your family" vs "your community" vs "your country" on self-reported intentions to wear a face covering. Results show that focusing on "your community" promotes intentions to wear a face covering relative to the baseline; the trend is the same when comparing "your community" to the other conditions, but not significant. We also conducted pre-registered analyses of gender differences on intentions to wear a face covering. We find that men less than women intend to wear a face covering, but this difference almost disappears in counties where wearing a face covering is mandatory. We also find that men less than women believe that they will be seriously affected by the coronavirus, and this partly mediates gender differences in intentions to wear a face covering (this is particularly ironic because official statistics actually show that men are affected by the COVID-19 more seriously than women). Finally, we also find gender differences in self-reported negative emotions felt when wearing a face covering. Men more than women agree that wearing a face covering is shameful, not cool, a sign of weakness, and a stigma; and these gender differences also mediate gender differences in intentions to wear a face covering.

**Introduction**

The coronavirus disease (COVID-19) pandemic represents a serious threat for millions of people around the world. In Bergamo the excess deaths in April 2020 (defined as the number of deaths in April 2020 minus the average number of deaths in the months before the COVID-19 outbreak) was equal to 4.5 times the baseline number of deaths; in Guayas, Ecuador, it was equal to 3.5 times the baseline number of deaths; in New York City it was equal to 3 times (Burn-Murdoch, Romei, & Giles, 2020). These numbers clearly tell about the deadly power of this coronavirus: since the baseline number of deaths worldwide is about 57 million people a year, if the coronavirus were to hit the planet with the same power as it hit Bergamo, Guayas or New York, somewhere between 150 millions and 250 millions people could die in a year.

Critically, a great deal of these deaths are not due directly to the coronavirus, but to the fact that hospitals get overwhelmed and people cannot be treated as they should (Fink, 2020; Glanz et al., 2020). For this reason, dozens of countries have imposed restrictions to the free circulation of people, with the short-term goal of slowing down the exponential growth of the virus and "flatten the curve" of the spread. At the time of this writing, about one third of the world population is under some form of restriction (Kaplan, Frias, & McFall-Johnsen, 2020). However, restrictions are highly costly in the long-term, both economically and psychologically (Van Bavel et al. 2020). Therefore, the countries in which the curve of the spread has been flattened (at least to some extent) are preparing to (or have already) lift(ed) some of the restrictions; these include some of the most severely hit countries, such as Italy, Spain, and the USA.

Yet, since a cure to the COVID-19 has not been found yet, it is of utmost importance that, when shelter-in-place rules are relaxed, people take preventive measures to avoid that the virus starts exponentially spreading again. One of the key prevention measures that have been discussed by medical researchers and policy makers is wearing a face covering, which has even been made mandatory in several European countries as well as in many American counties, while it is strongly suggested in others (Javid, 2020).

*Messaging to promote intentions to wear a face covering*

To adhere to a rule such as wearing a face covering is difficult because it requires a substantial change in our habits. Therefore, the risk that people do not adhere to it is high. For this reason, finding mechanisms that can promote the use of face covering is key during this phase of the pandemic response. Regulations that explicitly punish the violation of the rule are certainly crucial to impose behavioral changes. However, since it is impossible to control each and every person in a country, it is important to complement explicit governmental laws with implicit behavioral interventions designed to impact people's behavior without the force of law.

In particular, messages highlighting the costs (or the benefits) of (not) failing to respect prevention measures can be very effective, as they can be displayed almost everywhere in the street through screens and posters; they can reach people in their homes through television and social media; and they can even be voiced in the street using cars equipped with a megaphone, as it happened in Italy (Provantini & Ugolini, 2020). This raises an important question. Which types of messages are most effective in promoting pandemic responses that are in line with the recommendations of the medical profession? Social and behavioral science can be helpful in answering this question (Van Bavel et al. 2020). Accordingly, several works in the past month have explored the effect of several types of messages on pandemic response (Bilancini et al. 2020; Everett et al. 2020; Heffner, Vives, & FeldmanHall, 2020; Jordan et al. 2020).

From the viewpoint of classical economic theory, people are inherently self-regarding, which suggests that messages focusing on the consequences at the individual level might be more effective than messages focusing on the consequences on others, even genetically related others. Yet, decades of social and psychological research shows that people sometimes do not act solely to maximize their own payoff: at least in economic games, a substantial proportion of people appear to be driven by moral preferences (Krupka & Weber, 2013; Kimbrough & Vostroknutov, 2016; Eriksson et al., 2017; Capraro & Rand, 2018; Tappin & Capraro, 2018). Research using electric shocks has also shown that people tend to weigh harm to others more than harm to self (Crockett et al. 2014). This suggests that highlighting the consequences on other people may be a more effective strategy than highlighting the consequences to the self.

Consistent with this view, Jordan et al. (2020) have recently reported an experiment on Amazon Mechanical Turk (with American participants) where they found that telling participants that the coronavirus is a threat to "their community" is more effective at increasing prevention intentions than telling participants that the coronavirus is a threat to "themselves". However, in their measure of preventive measures, no item regarding the use of face covering was included.

The first contribution of our work is that we extend Jordan et al.'s work both in terms of the set of messages and in terms of preventive measures. In terms of messages, we test the relative effectiveness of messages highlighting that the coronavirus is a threat for people's family and for people's compatriots, beyond the messages already used by Jordan et al. Second, we focus on intentions to engage in a type of preventive behavior that is particularly important now that shelter-in-place rule are being relaxed: wearing a face covering. This preventive behavior was not considered in previous research.

### *Gender differences in intentions to wear a face covering and in emotions felt when wearing a face covering*

The second contribution of our work regards a detailed (pre-registered) analysis of gender differences in self-reported intentions to wear a face covering and in the self-reported negative emotions felt when wearing a face covering. The rationale for this analysis comes from the observation that previous work has found that men intend to engage in preventive behaviors less than women do (Jordan et al. 2020). Therefore, we predicted that we would find gender differences also in intentions to wear a face covering. Clearly, understanding whether there are such differences could be important to tailor interventions specifically on men. For this reason, we reasoned that, beyond asking intentions to wear a face covering, it would have been important to also collect self-reported emotions felt when wearing a face covering. The idea is that it is possible that men more than women report having negative feelings when wearing a face covering, and this could eventually mediate the gender differences in intentions to wear a face covering.

After the first round of data collection (see Methods for details), we noticed another interesting pattern of results related to gender. It seemed that gender differences in intentions to wear a face covering were particularly strong in counties were wearing a face covering was

*not* mandatory, while they almost disappeared in counties were wearing a face covering was mandatory. In other words, it seemed that making mandatory the wear of a face covering affected men's intentions to a greater extent than women's intentions. Therefore, we took advantage of the second session of data collection to further explore this effect. Moreover, in the second session, we decided to include questions about the subjective likelihood to get the coronavirus disease and, if so, the subjective likelihood to get over it relatively easily. Our rationale was that it is possible that these variables mediate gender differences in intentions to wear a face covering, and maybe also in emotions felt when wearing a face covering. We knew that this would be particularly ironic because medical studies do show that men are affected by the COVID-19 more seriously that women (Cai, 2020; Chen et al. 2020).

**Method**

*Conditions*

Participants were randomly assigned to one of five conditions. In each case, they were shown a message. The key difference between the messages is that the message in the *You* condition stresses the fact that the coronavirus is a threat to "you" (i.e., the participant); the message in the *Your family* condition stresses the fact that the coronavirus is a threat to "your family"; the message in the *Your community* condition stresses the fact that the coronavirus is a threat to "your community"; the message in the *Your country* condition stresses the fact that the coronavirus is a threat to "your country". See Table 1 for the exact messages.

| Condition | Message |
| --- | --- |
| Baseline | Various regions of the US are or will soon be moving towards the second phase of the coronavirus response strategy. Shelter-in-place rules will be relaxed and as a consequence some segments of the population will be allowed to move around more freely. |
| You | Various regions of the US are or will soon be moving towards the second phase of the coronavirus response strategy. Shelter-in-place rules will be relaxed and as a consequence some segments of the population will be allowed to move around more freely. However, since a cure for the coronavirus (COVID-19) has not been found yet, the COVID-19 remains **a serious threat to you**. Fortunately, **there are steps you can take when you go out to keep you safe**, including wearing a face covering and practicing social distancing. |
| Your family | Various regions of the US are or will soon be moving towards the second phase of the coronavirus response strategy. Shelter-in-place rules will be relaxed and as a consequence some segments of the population will be allowed to move around more freely. However, since a cure for the coronavirus (COVID-19) has not been found yet, the COVID-19 remains **a serious threat to your family**. Fortunately, **there are steps you can take when you go out to keep your family safe**, including wearing a face covering and practicing social distancing. |
| Your community | Various regions of the US are or will soon be moving towards the second phase of the coronavirus response strategy. Shelter-in-place rules will be |

|  | relaxed and as a consequence some segments of the population will be allowed to move around more freely. However, since a cure for the coronavirus (COVID-19) has not been found yet, the COVID-19 remains **a serious threat to your community**. Fortunately, **there are steps you can take when you go out to keep your community safe**, including wearing a face covering and practicing social distancing. |
|---|---|
| Your country | Various regions of the US are or will soon be moving towards the second phase of the coronavirus response strategy. Shelter-in-place rules will be relaxed and as a consequence some segments of the population will be allowed to move around more freely. However, since a cure for the coronavirus (COVID-19) has not been found yet, the COVID-19 remains **a serious threat to the US**. Fortunately, **there are steps you can take when you go out to keep Americans safe**, including wearing a face covering and practicing social distancing. |

*Table 1. Conditions of the experiment. Between-subjects random assignment.*

### *Dependent variables*

After reading the message, all participants took three scales. The first two scales were taken in random order.

1) *Intentions to wear a face covering*. Participants were asked to "be as accurate and honest as [they] can" when rating the following items: When the shelter-in-place rules are relaxed, I intend to ...
    a. Wear a face covering any time I leave home.
    b. Wear a face covering any time I am engaged in essential activities and/or work, and there is no substitute for physical distancing and staying at home.
    c. Wear a face covering any time I'm around people outside my household.

2) *Intentions to practice physical distancing*. Participants were asked to "be as accurate and honest as [they] can", when rating the following items: When the shelter-in-place rules are relaxed, I intend to ...
    a. Avoid hugging.
    b. Avoid kissing.
    c. Keep physical distance.
    d. Avoid handshaking.

After the main scales, participants took a third scale.

3) *Emotion felt from wearing a face covering*. Participants were asked to what extent they agree with the following statements:
    a. Wearing a face covering is cool.
    b. Wearing a face covering is not cool.
    c. Wearing a face covering is shameful
    d. Wearing a face covering is a sign of weakness.

e. The stigma attached to wearing a face covering is preventing me from wearing one as often as I should

All answers were collected using a 10-line "snap to grid" slider with three labels: "strongly disagree" at the far left, "neither agree nor disagree" at the center, "strongly agree" at the far right.

### *Demographics*

After the scales, participants were asked a set of demographic variables: sex, age, race, political views, religiosity, whether they live in an urban area, whether wearing a face covering is mandatory in their county, and whether there live in an area where shelter-in-place rules apply. There were also some other demographics, but these depended on the experimental session. In Session 1 we asked participants whether they were tested positive, whether they were tested negative, whether they were not tested but believe to have had the coronavirus. In session 2, we replaced the last two questions with two questions regarding the subjective likelihood of getting infected and the subjective likelihood of recovering relatively easily in case they get infected. These variables were formulated as follows. We asked subjects to report the extent to which they agree with the statement "I believe that it is unlikely that I will get the coronavirus (COVID-19)" and with the statement "If I get the coronavirus (COVID-19), I believe I will get over it relatively easily". Answers were collected using a 7-point likert scale from 1 = "strongly disagree" to 7 = "strongly agree". At the end, there was one control question to get rid of potential bots.

### *Pre-registration*

The design[1] and the analyses of Session 1 were pre-registered at: https://aspredicted.org/xj837.pdf. Those of Session 2 were pre-registered at: https://aspredicted.org/yr6p4.pdf. As pre-registered in Session 2, the reason for conducting two sessions is that Session 1 gave some inconclusive results that we wanted to test on a larger sample. Here we report the analysis directly on the overall sample. We conducted analysis with and without a dummy variable taking into account the experimental session. All the results remain qualitatively the same. Also, in the main text we focus on the results relative to wearing a face covering; the results on physical distancing were uninteresting and so we relegate them to the Appendix. We conduct the relevant analyses with and without a dummy variable that takes into account which scale was taken first. All the results remain qualitatively the same. Therefore, in the analysis that follows we report the outcomes without these dummy variables.

## Results

### *Participants*

The experiment was conducted in two sessions, the first one on April 28, 2020, the second one on May 4. In total, we recruited 2,516 participants living in the USA using Amazon Mechanical Turk (Paolacci, Chandler, & Ipeirotis, 2010). Those who participated in session 1 (as identified by their Turk ID) were not allowed to participate in Session 2. Within the same session, we found 32 multiple IP addresses and multiple Turk IDs; for each of them, we kept only the first observation. Additionally, we found that 25 subjects did not pass the bot test. As pre-registered, we eliminated these subjects. Thus, we remained with 2,459 subjects. Table 2 reports the demographic characteristics of the overall sample. The sample is quite heterogeneous, although not representative: males are slightly overrepresented; the age group 25-54 is overrepresented, at the cost of the age groups 18-24 and 65+, which are underrepresented; whites are also overrepresented, while Blacks or African Americans are underrepresented.[2] Moreover, the average participant appears to be also more left-leaning than the average American.[3] The sample is representative of people living in urban areas.[4] We could not find the exact percentage of counties where wearing a face covering was mandatory, neither we could find the exact percentage of counties where there were shelter-in-place rules; in the Table we report also these proportions for completeness.

---

[1] In the pre-registration, we forgot to include the way the scales were randomized.

| **Demographic** | | **Frequency** | **Percent** |
|---|---|---|---|
| gender | female | 1183 | 48.10 |
| | male | 1266 | 51.48 |
| | prefer not to say | 10 | <0.01 |
| age | 18-24 | 146 | 5.93 |
| | 25-34 | 838 | 34.08 |
| | 35-44 | 644 | 26.19 |
| | 45-54 | 423 | 17.20 |
| | 55-64 | 276 | 11.22 |
| | 65+ | 131 | 5.32 |
| race | American Indian or Alaska native | 14 | 0.56 |
| | Asian | 199 | 8.09 |
| | Black or African American | 143 | 5.81 |
| | Native Hawaiian or other Pacific Islander | 7 | <0.01 |
| | White | 2025 | 82.35 |
| | Multiracial | 70 | 2.85 |
| political view | left-leaning | 1241 | 50.46 |
| | center | 507 | 20.61 |
| | right-leaning | 708 | 28.79 |
| living in an urban or suburban area | | 1909 | 77.63 |
| living in a county where wearing a face covering is mandatory | | 1069 | 43.47 |
| living in a county where there are shelter in place rules | | 2129 | 86.58 |

*Table 2. Demographic characteristics of the overall sample. Political view goes from 1 = "very left-leaning" to 7 = "very right-leaning", with 4 = "center". In the Table we classified as "center" only those subjects who answered "center". In some cases, the percentages do not sum up to 100% because of some missing data. For example, in the political view question there are three missing observations.*

---

[2] https://en.wikipedia.org/wiki/Demographics_of_the_United_States
[3] https://news.gallup.com/poll/275792/remained-center-right-ideologically-2019.aspx
[4] https://www.census.gov/newsroom/press-releases/2016/cb16-210.html

*The effect of messages on self-reported intentions to wear a face covering*

We begin by building the composite variable *intentions to wear a face covering* by taking the average of its three items ($\alpha = 0.932$). Figure 2 reports the mean value of this variable, split by condition (error bars represent standard error of the mean). As pre-registered, we make pairwise comparisons using Wilcoxon rank-sum to test for differences across conditions. We find that the *Your community* condition gives rise to greater intentions to wear masks than the *Baseline* (p = 0.021) and, marginally, than the *Your family* condition (p = 0.065). All other p's > 0.1.

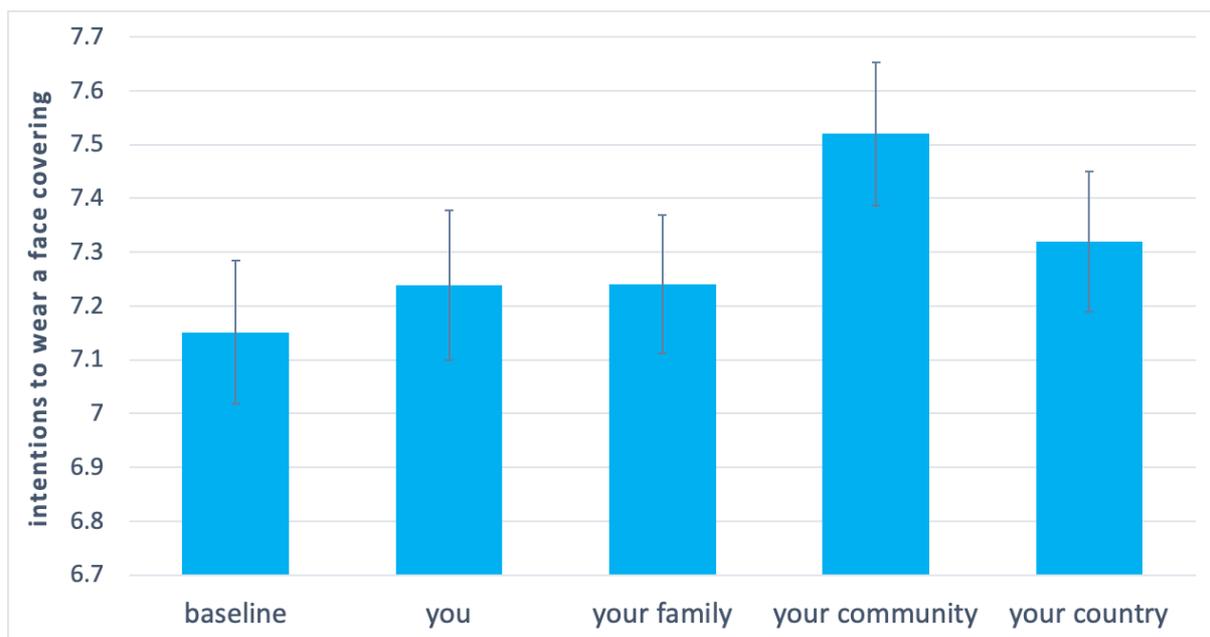

*Figure 1. Mean values of the "intention to wear a face covering" variable, split by condition. Error bars represent standard errors of the means.*

We also conduct exploratory analysis using linear regression to test whether this effect is robust to the inclusion of demographic variables. We find that the difference between the *Your community* condition and the *Baseline* remains significant when including all controls (b = 0.404, p = 0.020). All other comparisons are not significant (all p's > 0.1). Moreover, we find than none of the main demographic variables moderate the effect (gender: p = 0.638; age: p = 0.490; race: p = 0.885; religion: p = 0.492; urban: p = 0.885; face covering mandatory: p = 0.926; shelter-in-place rules: p = 0.995). The only individual characteristics that moderated the effect was political orientation (b = 0.252, p = 0.017). The positive sign suggests that the positive effect of focusing on your community was driven by people who self-report being right-leaning. Indeed, exploratory analysis shows that the effect is absent among people who self-report being left-leaning (b = 0.115, p = 0.497) and it is marginally significant among subjects who self-report being right-leaning (b = 0.722, p = 0.075).

*Individual differences in self-reported negative emotions felt when wearing a face covering*

In the first session, we pre-registered that we would test for the effect of gender and age on the "negative emotions felt wearing a face covering". Here, we report this analysis directly on the whole sample (session 1 and session 2 together). First, we build this composite variable by taking the average of its five items, after reversing the first item ($\alpha = 0.747$). Then, as pre-registered, we use linear regression to check the effect of sex and age on this variable, first in the baseline condition and then in all conditions together, in order to test for robustness. In the baseline, we find that being a female is significantly less associated with negative emotions felt wearing a face covering (b = -0.329, p = 0.037), as it is also being older (b = -0.012, p = 0.049). Putting all conditions together, we obtain a similar effect of gender (b = -0.398, p < 0.001), while the effect of age becomes smaller and marginally significant (b = -0.004, p = 0.059). If we control, as pre-registered, for all the demographic variable (see Table 3), both the effects of gender and age are highly significant. Controlling for the demographic variables also reveal a significant effect of political views (right-leaning people tend to have more negative feelings when wearing a face covering). We also conduct non-preregistered exploratory analyses over the "intentions to wear a face covering" variable to see whether these individual differences remain significant. We find that they do. Being a female is associated with greater intentions to wear a face covering; the same holds true for being left-leaning and for being older. Interestingly, we find that living in a county where wearing a face covering is mandatory impacts people's intentions to wear a face covering, but not the negative emotions that they feel when wearing a face covering. We refer to Table 3 for regression details.

| VARIABLES | Intentions to wear a face covering | Negative emotions felt wearing a face covering |
| --- | --- | --- |
| female | 0.462*** | -0.328** |
|  | (0.111) | (0.071) |
| age | 0.011*** | -0.008*** |
|  | (0.004) | (0.002) |
| Asian | 1.694** | 0.312 |
|  | (0.756) | (0.486) |
| Black or African-American | 1.557** | 0.274 |
|  | (0.768) | (0.493) |
| Native Hawaiian or other Pacific Islander | 2.073 | -0.760 |
|  | (1.267) | (0.814) |
| White | 0.899 | 0.169 |
|  | (0.732) | (0.470) |
| Multiracial | 1.547* | -0.592 |
|  | (0.801) | (0.514) |
| Right-leaning | -0.389*** | 0.267*** |
|  | (0.036) | (0.023) |
| Religion | 0.003 | -0.010 |
|  | (0.015) | (0.010) |
| Living urban area | -0.163 | 0.125 |
|  | (0.133) | (0.085) |
| Face covering not | 1.371*** | -0.090 |
| mandatory in county | | |
| | (0.115) | (0.074) |
| Shelter-in-place rules active in county | 0.354** | -0.191* |
| | (0.167) | (0.107) |
| Constant | 5.757*** | 3.04*** |
| | (0.798) | (0.512) |
| | | |
| Observations | 2,453 | 2,453 |
| R-squared | 0.145 | 0.083 |

*Table 3. Regression details. Standard errors in parentheses. \*\*\* p<0.01, \*\* p<0.05, \* p<0.1*

### Gender differences as a function of whether wearing a face covering is mandatory

As pre-registered, we now explore more in-depth gender differences in intentions to wear a face covering and in negative emotions felt when wearing a face covering, as a function of whether wearing a face covering is mandatory in the county where the participant lives. For this analysis, we exclude 10 subjects who responded that they prefer not to say their gender. Figure 3 summarizes the results. Linear regression predicting intentions to wear a face covering as a function of sex, a dummy variable that takes into account whether the face covering is mandatory, and their interaction, reveals a significant main effect of gender (b = 0.776, p < .001), a significant main effect of whether wearing a face covering is mandatory (b = 1.755, p < .001), and, crucially, a significant interaction (b = -0.499, p = 0.031). The interaction is driven by the fact that gender differences in intentions to wear a face covering are very strong when wearing a face covering is not mandatory (b = 0.720, p < .001), but less so when wearing a face covering is mandatory (b = 0.298, p = 0.026). All these effects remain qualitatively the same when we include controls on all the other demographics.

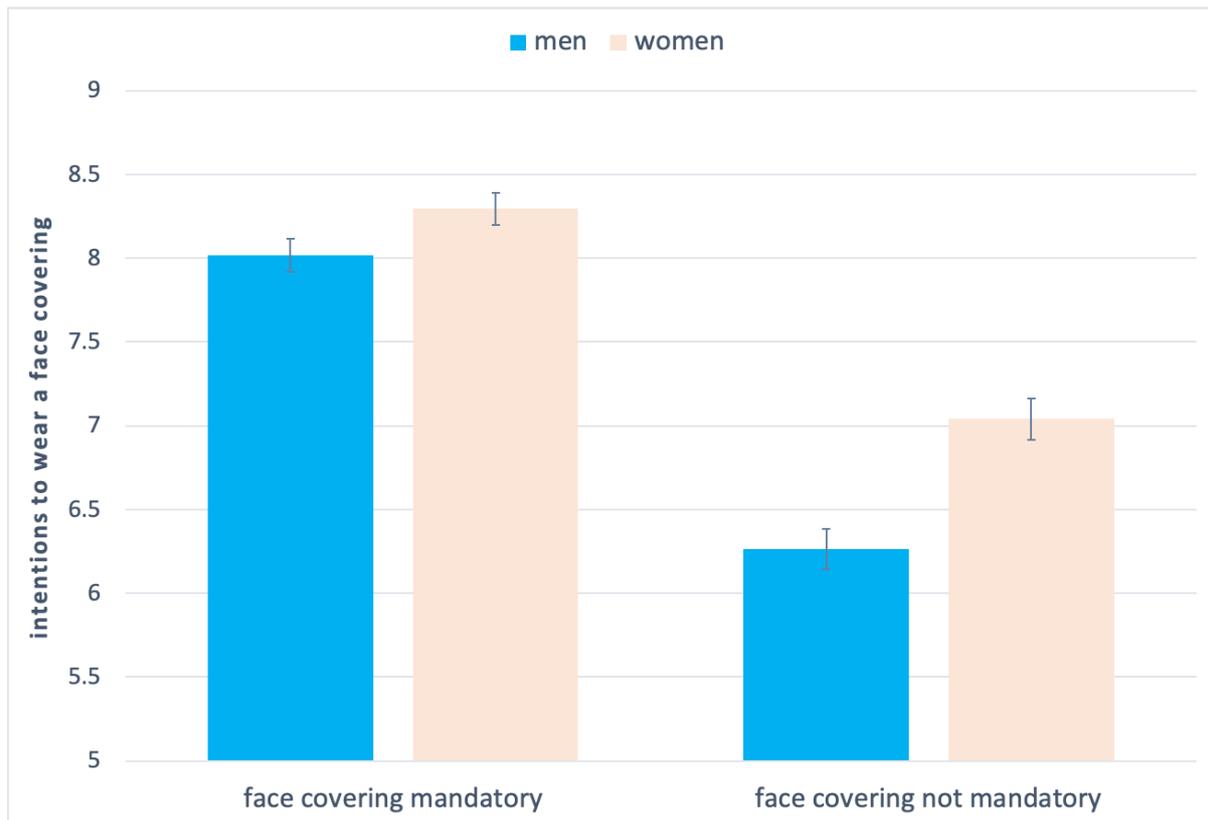

*Figure 2. Gender differences in intentions to wear a face covering, split by whether wearing a face covering is mandatory in the county where the participant lives.*

Then we look at gender differences in negative emotions felt when wearing a face covering. For this analysis too, we exclude 10 subjects who responded that they prefer not to say their gender. Figure 4 summarizes the results. Linear regression predicting the negative emotions felt when wearing a face covering as a function of sex, whether the face covering is mandatory, and their interaction, reveals a significant main effect of gender (b = -0.435, p < .001), a significant main effect of whether wearing a face covering is mandatory (b = -0.204, p =0.048), and a statistically non-significant interaction (b = 0.067, p = 0.653). Note, however, that, as we have seen before, the main effect of whether wearing a face covering is mandatory loses significance when controlling for all other variables; and this happens also when we include the interaction between sex and whether wearing the face covering is mandatory, in which case the main effect of whether wearing the face covering is mandatory is not significant (p=0.187).

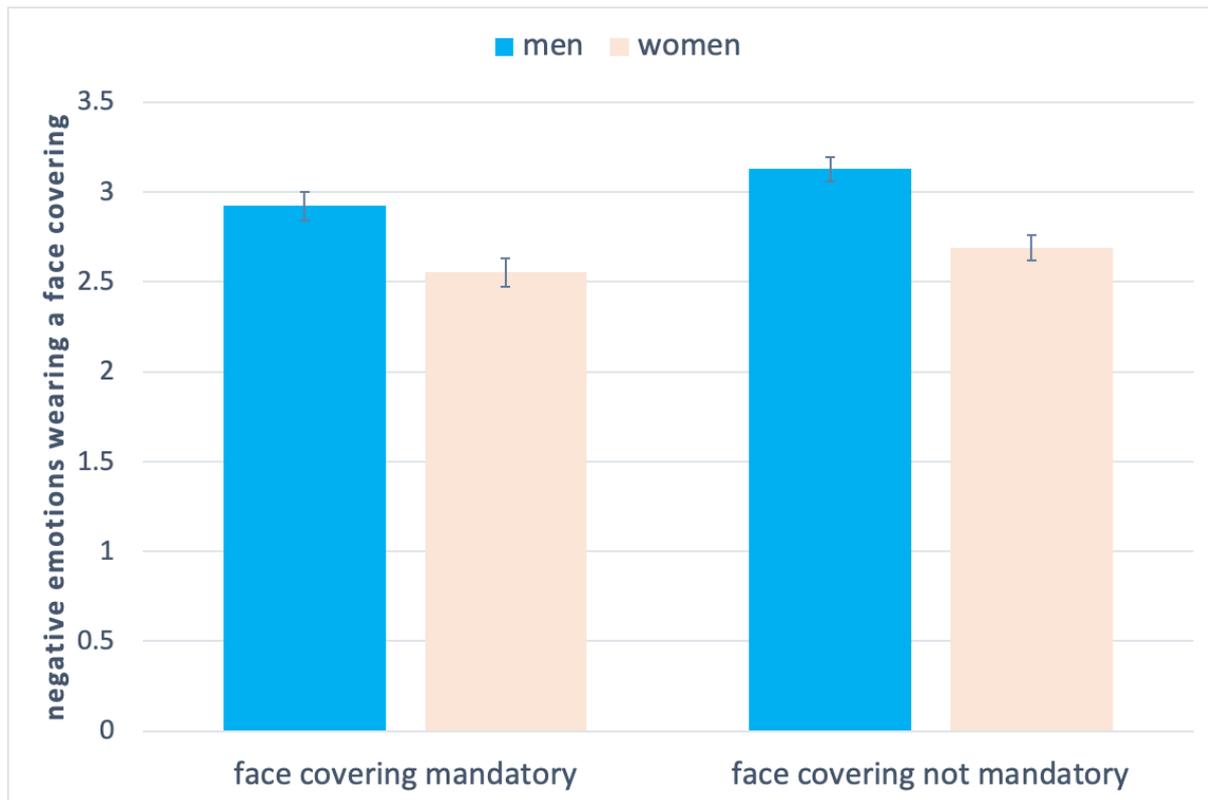

*Figure 3. Gender differences in negative emotions felt when wearing a face covering, split by whether wearing a face covering is mandatory in the county where the participant lives.*

*Mediation analyses*

In Session 2 we pre-registered that we would test: (i) whether gender differences in intentions to wear a face covering are mediated by the negative emotions felt when wearing a face covering, (ii) whether gender differences in intentions to wear a face covering are mediated by the perceived likelihood of getting infected and, if so, by the perceived likelihood to recover from the infection relatively easily, and (iii) whether gender differences in negative emotions felt when wearing a face covering are mediated by the perceived likelihood of getting infected and, if so, by the perceived likelihood to recover from the infection relatively easily.

Starting from the first mediation analysis, linear regression predicting intentions to wear a face covering as a function of gender and emotions felt when wearing a face covering reveals that the negative emotions felt from wearing a face covering have a significant negative effect on intentions to wear a face covering ($b = -0.686$, $p < .001$) and that the coefficient of the gender variable ($b = 0.317$, $p = 0.003$) is smaller than the coefficient of the gender variable that is obtained when regressing intentions to wear a face covering over gender only ($b = 0.590$, $p < .001$). This suggests that, indeed, negative emotions felt when wearing a face covering partly mediates gender differences in intentions to wear a face covering.

We now pass to the other two mediation analyses. First of all, we test whether there are gender differences in the two measures that are supposed to mediate the effect. Indeed, linear regression finds that men more than women believe that it is *unlikely* that they will get infected ($b=0.318$, $p <.001$); in a scale from 0 to 6, the mean values are 3.25 (SE = 0.06) for men and 2.94 (SE = 0.06) for women. Similarly, men more than women believe that, in case they get infected, they will get over it relatively easily (b = 0.217, p = 0.009); in a scale from 0 to 6, the mean values are 3.42 (SE = 0.06) for men and 3.20 (SE = 0.06) for women.

So, we pass to the mediation analyses to see whether these variables mediate gender differences in intentions to wear a face covering and in the negative emotions felt when wearing a face covering.

Regressing the intentions to wear a face covering over gender and the subjective likelihood to get infected, we find that the subjective probability to get infected is significant (b = -0.417, p < .001) and that the coefficient of the gender variable (b = 0.462) is smaller than the coefficient of the gender variable that is obtained when regressing the intentions to wear a face covering over gender only (b = 0.590). This indicates that, indeed, the subjective likelihood to get infected partly mediates gender differences in intentions to wear a face covering. A qualitatively similar result is obtained replacing the "subjective likelihood to get infected" variable with the "subjective likelihood to recover in case one gets infected" variable: when regressing intentions to wear a face covering over gender and the likelihood to get over it if infected, we find that the perceived likelihood variable is significant (b = -0.577, p < .001) and that the coefficient of the gender variable (b = 0.471) is smaller than the coefficient of the gender variable that is obtained when regressing the intentions to wear a face covering variable over the gender variable only (b = 0.590).

Interestingly, we find that the same variables, the subjective likelihood to get infected and the subjective likelihood to get over it in case one gets infected, do *not* mediate the gender difference in the negative emotions felt when wearing a face covering. Regressing the negative emotions felt when wearing a face covering over gender and the subjective likelihood to get infected, we do find that the likelihood to get infected variable is significant (b = 0.145, p < .001), but this time the coefficient of the gender variable (b = -0.396) is essentially the same as the coefficient of the gender variable that is obtained when regressing the negative emotions felt when wearing a face covering over the gender variable only (b = -0.398). Similarly, regressing the negative emotions felt when wearing a face covering over gender and the subjective likelihood to get over the disease easily in case one gets it, we do find that the likelihood to get over it variable is significant (b = 0.145, p < .001), but this time the coefficient of the gender variable (b = -0.400) is essentially the same as the coefficient of the gender variable obtained when regressing the negative emotions felt when wearing a face covering over the gender variable only (b = -0.398). In sum, gender differences in negative emotions felt from wearing the face covering do not seem to be explained by gender differences in the subjective likelihood of getting infected and, in case so, of getting over it easily; on the other hand, gender differences in these two variables partly explain gender differences in intentions to wear a face covering.

*Exploratory analysis*

As additional exploratory analysis, we would like to better understand the reasons why men tend to feel stronger negative emotions when wearing a face covering. Understanding this might help think about particular interventions focused to promote the use of face covering among men. To this end, we look at gender differences at the item level. We find gender differences in all items: when people are asked whether they agree with the statement "wearing a face covering is cool" (b = 0.249, p = 0.032), when people are asked whether they agree with the statement "wearing a face covering is not cool" (b = -0.363, p = 0.006), when people are asked whether they agree with the statement "wearing a face covering is shameful" (b = -0.472, p < .001), when people are asked whether they agree with the statement "wearing a face covering is a sign of weakness" (b = -0.481, p < .001), and when people are asked whether they agree with the statement "the stigma attached to wearing a face covering is preventing me from wearing one as often as I should" (b = -0.489, p < .001). These results are robust to the inclusion of all the other demographic controls.

**Discussion**

Now that several countries are moving towards relaxing shelter-in-place rules, it is important that people engage in preventive behaviors, such as wearing a face covering, to avoid that the coronavirus disease (COVID-19) starts exponentially spreading again. To promote adherence to this rule, laws or mandates should be supplemented with behavioral interventions.

Here, we reported an online experiment with a heterogeneous, although not representative, large sample of people living in the USA, where we tested the relative effect of messages highlighting that the coronavirus is a threat to "you" vs "your family" vs "your community" vs "your country" on self-reported intentions to wear a face covering. Results show that focusing on "your community" is better than the baseline; we also find a common trend such that focusing on "your community" seems to be slightly more effective than focusing on "you", "your family" and "your country", but none of the pairwise comparisons were statistically significant (only the one vs. "your family" was marginally significant).

These results are similar to those presented by Jordan et al. (2020), although weaker. Specifically, Jordan et al. found that focusing on "your community" promotes intentions to engage in preventive behaviors compared to the "baseline" (as we also do) and compared to focusing on "you" (in our case we find a non-significant trend). Our design differs from Jordan et al.'s in two regards. First, in their experiment, participants do not only read a message but are also shown a flier, which might have contributed to increase the size of the effect. The second difference regards the dependent variable: Jordan et al. (2020) use a set of prevention measures (handwashing, avoid touching one's own face, etc.) that is disjoint from our set of measures. Therefore, it is also possible that Jordan et al.'s results simply do not extend to our measure. In any case, putting together our results and those of Jordan et al. (2020) we can conclude that focusing on "your community" promotes intentions to engage in several preventive behaviors compared to the baseline. This can be a useful recommendation for leaders and policy makers.

Our pre-registered analysis of gender differences revealed a number of interesting results. Men less than women intend to wear a face covering. This is true especially in counties where wearing a face covering is not mandatory. Indeed, in counties in which wearing the face covering is mandatory, gender differences in intentions to wear a face covering almost disappear. This suggests that making wearing a face covering mandatory has a larger effect on men than on women. Moreover, we found that gender differences in intentions to wear a face covering are mediated by the subjective likelihood to get the disease and by the subjective likelihood to get over it more easily than not, in case one gets it. In other words, the fact that men less than women intend to wear a face covering can be partly explained by the fact that men more than women believe that they will be relatively unaffected by the disease. This is particularly ironic because official statistics show that actually the coronavirus (COVID-19) impacts men more seriously than women. For example, 60% of the deaths are men (Cai, 2020; Chen et al. 2020).

We also found that more men than women tend to report negative emotions when wearing a face covering. Moreover, negative emotions when wearing a face covering mediates gender differences in the intentions to wear a face covering. However, interestingly, gender differences in negative emotions felt when wearing a face covering does not seem to depend on whether wearing a face covering is mandatory. In other words, making the wear of a face covering mandatory changes the self-reported intentions to wear a face covering, but not the self-reported emotions felt when wearing it. Moreover, we found that the self-reported negative emotions felt when wearing a face covering were not mediated by the likelihood to get the disease and by the likelihood to get over it easily in case one gets it. We also conducted some exploratory analyses to test in which specific items of the "negative emotions felt when wearing a face covering" scale the gender differences were concentrated. We found that men more than women disagree with the statement "wearing a face covering is cool" and agree with the statements: "wearing a face covering is not cool" "wearing a face covering is shameful", "wearing a face covering is sign of weakness", and "the stigma attached to wearing a face covering is preventing [them] from wearing one as often as [they] should". This suggests that future interventions to promote the use of a face covering among men can try to act on decreasing these negative emotions. More generally, future work can test whether priming the use of reason vs emotion (Levine et al. 2018; Capraro, Everett & Earp, 2019; Caviola & Capraro, 2020) can be effective at increasing intentions to use face covering among men.

We conclude with a theoretical observation. Apart from the baseline and the "you" condition, all other conditions (your family, your community, your country) use moral messages that are based on a combination of the harm/care dimension (threat) and the ingroup/loyalty dimension (your family, your community, your country) of morality (Graham et al. 2013; Haidt, 2012; Haidt & Joseph, 2004). While these are important dimensions of morality, they are not the only ones (Graham et al. 2013; Haidt, 2012; Haidt & Joseph, 2004; Curry, 2016; Curry et al. 2019). Future research could test moral messages tapping other dimensions of morality.

**Appendix. Analysis of the intentions to practice physical distance**

We build the composite variable *intentions to practice physical distancing* by taking the average of its four items ($\alpha = 0.902$). Figure A1 reports the mean value of this variable, split by condition (error bars represent standard error of the mean). As pre-registered, we make pairwise comparisons using Wilcoxon rank-sum to test for differences across conditions. We find no significant differences (all p's > 0.3).

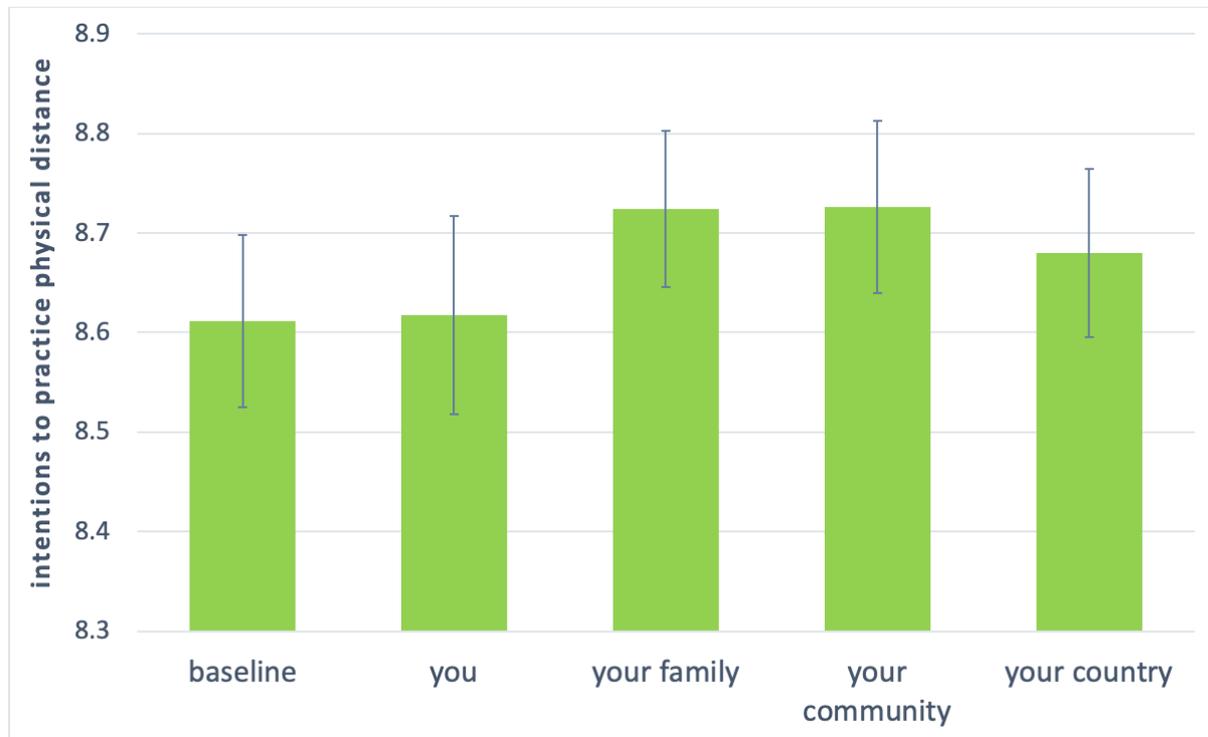

*Figure 3. Mean values of the "intentions to practice physical distancing" variable, split by condition. Error bars represent standard errors of the means.*

For completeness, we also report the full regression.

| VARIABLES | Intentions to practice physical distancing |
|---|---|
| female | 0.333*** |
|  | (0.076) |
| age | 0.014*** |
|  | (0.002) |
| Asian | 0.081 |
|  | (0.518) |
| Black or African-American | 0.710 |
|  | (0.525) |
| Native Hawaiian or other Pacific Islander | 0.433 |
|  | (0.866) |
| White | 0.052 |
|  | (0.501) |
| Multiracial | -0.010* |
|  | (0.548) |
| Right-leaning | -0.216*** |
|  | (0.024) |
| Religion | -0.005 |
|  | (0.010) |
| Living urban area | -0.116 |
|  | (0.091) |
| Face covering not mandatory in county | 0.107 |
|  | (0.079) |
| Shelter-in-place rules active in county | -0.385*** |
|  | (0.115) |
| Constant | 8.111*** |
|  | (0.546) |
|  |  |
| Observations | 2,453 |
| R-squared | 0.069 |

*Table A1. Regression details. Standard errors in parentheses. \*\*\* p<0.01, \*\* p<0.05, \* p<0.1*